\documentclass[runningheads]{llncs}
\usepackage[T1]{fontenc}
\usepackage{graphicx}
\usepackage{amsmath,cite,url}
\usepackage{graphicx}
\usepackage{color}
\usepackage{booktabs}
\usepackage{dsfont}
\usepackage{stfloats}
\usepackage{adjustbox}
\usepackage{siunitx}
\usepackage{multirow}
\usepackage[bookmarks=false]{hyperref}
\usepackage{bm}
\begin{document}
\title{GuitarFlow: Realistic Electric Guitar Synthesis From Tablatures via Flow Matching and Style Transfer}
\titlerunning{GuitarFlow}
\author{Jackson Loth\inst{1}\orcidID{0009-0002-8709-1218} \and
Pedro Sarmento\inst{1,2}\orcidID{0000-0002-4518-0194} \and
Mark Sandler\inst{1}\orcidID{0000-0002-5691-8107} \and
Mathieu Barthet\inst{1,3}\orcidID{0000-0002-9869-1668
}}
\authorrunning{J. Loth et al.}
\institute{Centre for Digital Music, Queen Mary University of London \email{\{j.j.loth,p.p.sarmento,mark.sandler,m.barthet\}@qmul.ac.uk} \and
Music.AI \and
Aix-Marseille Univ CNRS PRISM
}
%Springer Heidelberg, Tiergartenstr. 17, 69121 Heidelberg, Germany
%ABC Institute, Rupert-Karls-University Heidelberg, Heidelberg, %Germany\\
\maketitle              % typeset the header of the contribution

\begin{abstract} 
Music generation in the audio domain using artificial intelligence (AI) has witnessed steady progress in recent years. However for some instruments, particularly the guitar, controllable instrument synthesis remains limited in expressivity. We introduce GuitarFlow, a model designed specifically for electric guitar synthesis. The generative process is guided using tablatures, an ubiquitous and intuitive guitar-specific symbolic format. The tablature format easily represents guitar-specific playing techniques (e.g. bends, muted strings and legatos), which are more difficult to represent in other common music notation formats such as MIDI. Our model relies on an intermediary step of first rendering the tablature to audio using a simple sample-based virtual instrument, then performing style transfer using Flow Matching in order to transform the virtual instrument audio into more realistic sounding examples. This results in a model that is quick to train and to perform inference, requiring less than 6 hours of training data. We present the results of objective evaluation metrics, together with a listening test, in which we show significant improvement in the realism of the generated guitar audio from tablatures. 

\keywords{Flow matching  \and Style transfer \and Guitar \and  Synthesis \and Audio effects.}
\end{abstract}
\section{Introduction}
Recent advances in generative audio systems have given rise to impressive text-to-music systems that allow users to generate full songs from simple text prompts \cite{evans2024long} \cite{Forsgren_Martiros_2022}. While this is very interesting from a science and technology perspective, it is arguably lacking as a creative tool due to a lack of fine-grained control over the music missing. Music generation systems with full control over the notes do exist, but typically in the symbolic domain \cite{Sarmento2021dadagp}\cite{loth2023proggp}, requiring a way to transform a symbolic music score into audio. While this can be achieved through virtual instruments to varying degrees of success, guitars are particularly expressive instruments which are difficult to represent through the standard Musical Instrument Digital Interface (MIDI) \cite{loth2024midi}. Guitar timbre itself can even be perceived differently when performed in different playing styles \cite{loth_playing_2023}. Previous attempts at synthesising expressive music instruments have largely focused on MIDI or representations derived from MIDI \cite{wiggins2023differentiable}\cite{kim2024expressive}. % Removed \cite{hawthorne2022multi} because arXiv, and \cite{kim2024expressive} because space

%Our work builds on previous work in a few different ways. Previous works relied on diffusion, a popular generative technique, to synthesise audio. This is very computationally costly particularly when working with large dimensional data such as spectrograms. Flow matching has begun to be explored in the generative space as an alternative to diffusion that promises better performance for state of the art results \cite{esser2024scaling}, and is slowly starting to see use among audio applications \cite{wangfrieren} \cite{guan2024lafma}. It has also been shown to work effectively when synthesising latent representations of signals \cite{dao2023flow}. However, research is still in an early stage. 

This paper explores the task of synthesising electric guitar by processing synthetic audio rendered from an expressive symbolic musical representation. Instead of MIDI, we use guitar tablatures (see Figure \ref{fig:tabexample}) to represent the musical content of a desired audio synthesis. This allows us to incorporate expressive playing techniques such as slides, bends, hammer-ons, etc, something that other music instrument synthesis models struggle to do without additional conditioning mechanisms. We also introduce an intermediary step of first rendering a guitar tablature to audio using a quick and simple sample-based virtual instrument. Our model performs style transfer on this simple audio rendering and transforms it into more realistic sounding audio. We use `style` here to refer to various performance-related qualities which distinguish natural and more synthetic performances. The style transfer is achieved using Flow Matching \cite{lipman2022flow}, a recent generative modeling paradigm. This method allows us to greatly simplify the complexity of the training and inference pipeline, requiring less data and less time to train. 

The contributions of this paper are summarized as: (1) GuitarFlow, a novel model and methodology for realistic electric guitar synthesis from guitar tablatures using Flow Matching and style transfer; (2) an evaluation of the model using both objective metrics and a subjective listening test which shows the success of GuitarFlow in transforming audio rendered with a virtual instrument to sound more realistic; (3) a public repository\footnote{\href{https://github.com/JackJamesLoth/GuitarFlow}{https://github.com/JackJamesLoth/GuitarFlow}} of code to allow other researchers to replicate and extend the research. This paper demonstrates the potential of the technique in greatly lowering data and computational requirements when training generative audio models.
% In this paper, we first present some relevant background concerning guitar tablatures, style transfer and music instrument synthesis. In Section 3, we discuss the overall methodology for the experiment, including Flow Matching, and the model used. Section 4 details our experiments and we evaluate the model. Results are presented in Section 5 and the model and methodology, as well as their limitations, are discussed in Section 6.

%\textcolor{red}{In this paper, we first present some relevant background concerning previously released music datasets in symbolic format. In Section 3, we discuss advantages of tab-based datasets for MIR research. We then describe, in Section 4, the details of the DadaGP dataset, its encoder/decoder support tool, the features it encompasses and the ones it lacks. Within Section 5 we present a use case of symbolic music generation using our proposed dataset, supported by previous approaches concerning databases of symbolic music. Section 6 proposes additional applications for the dataset. Finally, in Section 7 we explain the steps needed in order to acquire the dataset, further pointing out some ethical considerations in Section 8. - FROM DADAGP}
\section{Background}

\subsection{Guitar Tablatures}

\begin{figure}
    \centering
    \includegraphics[width=0.7\columnwidth]{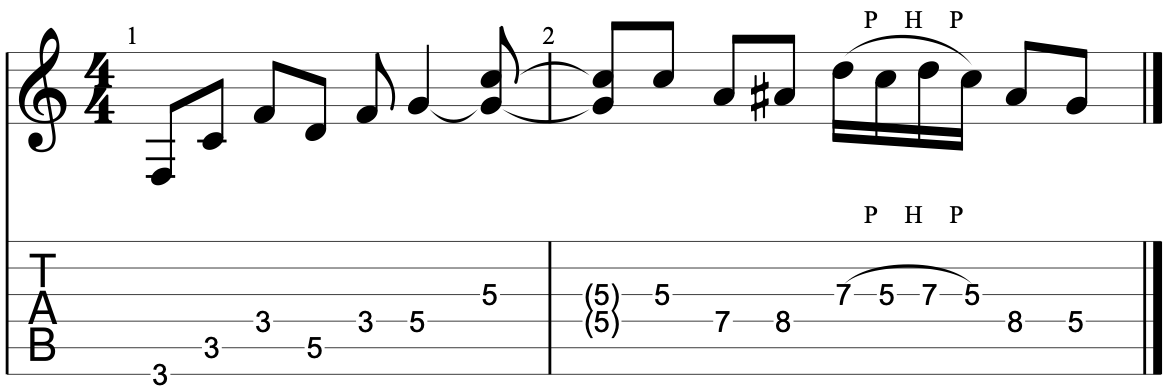}
    \caption{Example of a guitar tablature, from the Guitar Pro editing software.}
    \label{fig:tabexample}
\end{figure}

Guitar tablatures (refer to Figure \ref{fig:tabexample}), also known as tabs, are symbolic representations of guitar music and have seen increased attention in recent years due to their ability to easily represent guitar-specific expressions \cite{sarmento2024Guitarphd}. In contrast to MIDI, which simply represents a note's pitch and velocity over time, tabs represent both the fret and string number of a guitar. They can also support expressive playing techniques such as bends, hammer-ons, pull-offs, strum directions, and more. Tabs have seen an increase in attention from the MIR research community in the past few years in areas such as guitar tablature generation \cite{Sarmento2021dadagp} \cite{sarmento2023gtr}, automatic guitar transcription \cite{wiggins2019guitar} \cite{cwitkowitz2023fretnet} and tablature prediction from MIDI \cite{edwards2024midi}.

\subsection{Style Transfer}
% Removed [15] \cite{liu_unsupervised_nodate} \cite{zhang2023inversion}
% Removed 
Broadly speaking, style transfer involves transforming the ``style'' of a signal while preserving the ``content'' of signal. The task is popular in the image domain \cite{gatys_image_2016} \cite{zhu_unpaired_2017} . A common assumption is that the audio of an instrument can be broken up into ``content'', referring to the pitch, length and arguably loudness or intensity of the notes, and ``style'', referring to the actual sound of the instrument and expressiveness of the performer playing those notes. Some musical style transfer works use a GAN-based approach, training a generator to convert an audio input to the desired timbre \cite{huang_timbretron_2019}\cite{yang_unsupervised_2021}. Others use an autoencoder to produce disentangled representations of content and style \cite{mor_universal_2018} \cite{caillon_rave_2021}. As with generative tasks recently, Diffusion has also become a popular choice \cite{comanducci2023timbre} \cite{huang2024musical}, while differential digital signal processing (DDSP) \cite{engel_ddsp_2020} provides an alternative to all of these models by allowing trainable DSP functions.

\subsection{Music Instrument Synthesis}

Instruments such as guitar have traditionally been simulated by modeling the physics of the instrument and strings \cite{laurson2001methods}. More recently, models such as WaveNet \cite{oord_wavenet_2016} opened the door to neural generation conditioned on symbolic musical representations \cite{hawthorne2018enabling} \cite{kim2019neural}. DDSP has seen a lot of use creating audio synthesis models \cite{caspe2022ddx7} \cite{shier2023differentiable} due to its flexibility. Despite its large data requirements, diffusion has also become a popular method for neural audio generation. While much work is focused on generating full song mixes \cite{Forsgren_Martiros_2022} \cite{evans2024long}, synthesising symbolic musical representations such as MIDI allows much finer musical control over the synthesised audio. Hawthorne et al. \cite{hawthorne2022multi} trained a Transformer-based Diffusion model for multi instrument synthesis, which was in turn expanded on by Kim et al. \cite{kim2024expressive} for specifically acoustic guitar. While the models showed promising results, they were held back by lack of annotated data. Maman et al. \cite{maman2024multi} used automatic MIDI transcription systems to help address this problem while adding conditioning on instrument timbre. All of these methods have the drawback of requiring large amounts of data and necessitating significant computational resources to train. For example, the training in \cite{maman2024multi} took 350 hours over three powerful Nvidia A100 GPUs.

\section{Methodology}

\subsection{Flow Matching}

Flow Matching (FM) \cite{lipman2022flow} is a generative modeling paradigm which provides a way to efficiently train continuous normalizing flows (CNF). It has recently picked up interest among audio researchers \cite{guan2024lafma} \cite{wang2024frieren}. Intuitively, the idea is to learn the direction to move a point sampled from some data distribution over time in order to transform it into a sample from a different data distribution. More formally, we would like to transform a data distribution $p_0(x_0)$ to another data distribution $p_1(x_1)$ by learning a velocity field $v_t$ which points in the direction in which we would like to transform a point at a given time $t \in [0,1]$. Given a vector field $u_t$ which generates a target probability flow $p_t$, we learn a time-dependent flow which matches $p_t$. To do this, we consider the ordinary differential equation (ODE):
\begin{equation}
    dx_t = v_t(x_t)dt
    \label{eq:ode}
\end{equation}

Flow Matching attempts to learn this time-dependent vector field $v_t$ by minimising a conditional Flow Matching (CFM) loss \cite{tong2024improving}. This is generalized to arbitrary source and target distributions, making CFM an intuitive and powerful method for transforming data distributions. We adopt rectified flow \cite{liu2023flow} by simply sampling $x_0$ and $x_1$ from our data and calculating the interpolated point $x = (1-t)x_0 + tx_1$ and target velocity $u_t(x|z) = x_1 - x_0$. We then train a neural network $\theta$ by minimising the following:
\begin{equation}
    \mathcal{L} = ||v_\theta(t,x)-(x_1 - x_0)||^2_2
\end{equation}

Once we have learned our $v_\theta(x,t)$, we can use an ODE solver \cite{torchdiffeq} to approximate the solution to Equation \ref{eq:ode} over some set of discrete time steps to transform source data to the target distribution.

%Note that while a Gaussian distribution is commonly used as the source distribution $x_0$, it is not strictly required \cite{tong2024improving}. This makes Flow Matching an intuitive and potentially powerful method for transforming data distributions.

%\begin{figure*}
%    \centering
%    \includegraphics[width=1\textwidth]{figs/latent_transform.png}
%    \caption{An example of how latent points can move through time $t$ during the Flow Matching inference process. Two features are selected from a Music2Latent embedding and plotted. The mean ($\times$) and covariance (ellipse) are plotted for each distribution in order to better visualise them. A total of 100 discrete time steps are used.}
%    \label{fig:latent_transform}
%\end{figure*}

\subsection{Method}

Our approach focuses on modeling the direct input (DI) signal, the raw output of an electric guitar, rather than the amplified and distorted signal which is typically heard in recordings. This allows us to simplify the task and offload distortion processing to digital amplifier models \cite{wright_real-time_2020} \cite{chen2024towards}.

%An important aspect of our approach is the focus on synthesis of the direct input (DI) of an electric guitar. A DI is the raw signal from the output of a guitar pickup that would normally then be fed through a guitar amplifier. The DI is rarely ever heard by listeners and guitarists, as it is almost always run through a guitar amplifier in recordings, live performances, and casual playing. Digital guitar amplifier modeling \cite{wright_real-time_2020} \cite{chen2024towards} is extremely accurate and has a very low computational cost. Modeling just the DI allows us to focus on modeling the sound of a guitar rather than the characteristics of an amplifier, and to simplify the complexity of the model by offloading this part of the final guitar sound to a separate model.

Our training pipeline involves first rendering synthetic audio from a guitar tablature using a virtual instrument. This and corresponding real guitar DI audio are then encoded into a latent space using Music2Latent\cite{pasini2024music2latent}, a pretrained autoencoder. The audio is broken into four second chunks in order to keep the audio length consistent and keep the data size reasonably small. Our model then learns a mapping from a latent distribution $p_0(x_0)$ of synthetic guitar DI rendered by our virtual instrument to a latent distribution $p_1(x_1)$ of real guitar DI. We can then use $x_0 \sim p_0$ and $x_1 \sim p_1$, corresponding to the synthetic and real guitar DI respectively, for training. This makes the model effectively a one-to-one style transfer model, as the real and synthetic guitar recordings are required to contain exactly the same musical content (i.e. notes, timings, expressive techniques, etc.).

\subsection{Model}

\begin{figure*}[h]
    \centering
    \includegraphics[width=0.8\textwidth]{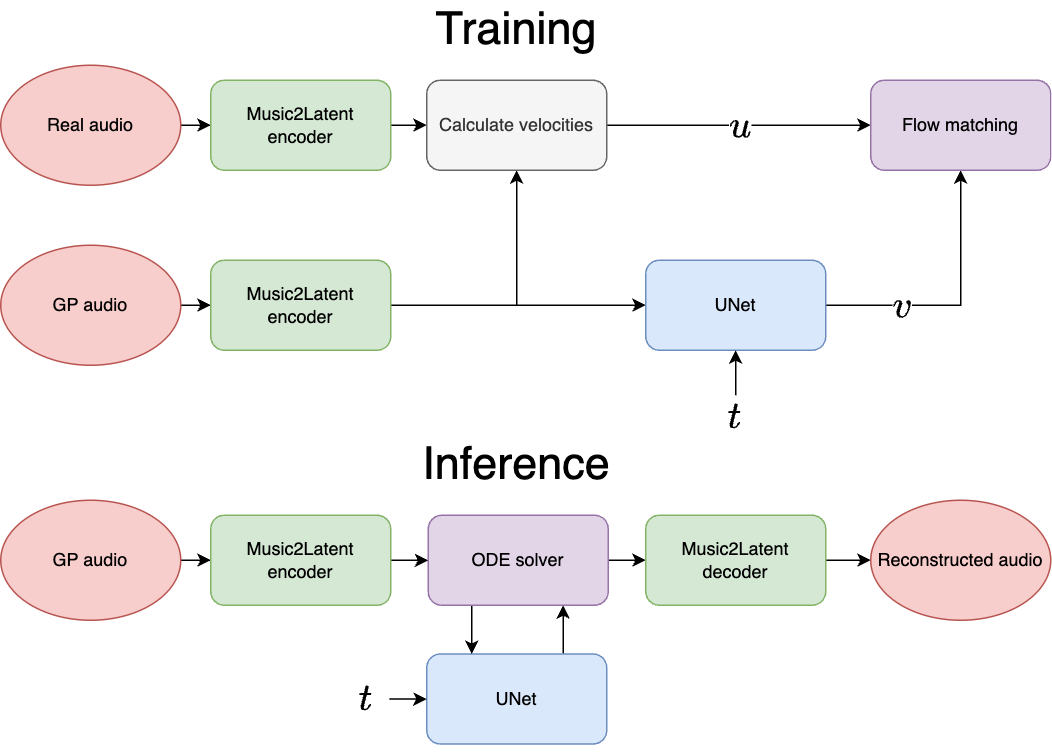}
    \caption{Training and inference using GuitarFlow.}
    \label{fig:model}
\end{figure*}

Our model, titled GuitarFlow, is primarily based on the UNet \cite{ronneberger_u-net_2015} architecture, consisting of four downsampling and upsampling layers\footnote{\href{https://github.com/clemkoa/u-net}{https://github.com/clemkoa/u-net}}. To condition the model on $t$, $x_0$ and $t$ are concatenated together in the feature level. The model then outputs predicted flow velocities $v_t$, which are used alongside real flow velocities $u_t$ to calculate mean squared error (MSE) loss. Figure \ref{fig:model} shows the training and inference pipeline.

For inference, an ODE solver \cite{torchdiffeq}  is used to integrate across the learned velocity field across 100 discrete time steps. We found the Dormand-Prince method \cite{dormand1980family} to yield significantly higher quality results compared to a more standard Euler ODE solver.
\section{Experiments}

\subsection{Data and Training}
The GOAT dataset \cite{loth2025goat} was used to train and evaluate our model. This dataset contained roughly 5.75 hours of real guitar DI audio recorded at 44.1kHz and the corresponding tablature annotations. The performances were performed by three guitarists on four different guitars, largely cover rock and indie rock genres and include a wide variety of expressive techniques. The guitar tablatures were rendered into audio using the Realistic Sound Engine (RSE) virtual instrument in Guitar Pro 7\footnote{\href{https://www.guitar-pro.com/blog/p/14545-signature-sounds-explained-guitar-pro-7}{https://www.guitar-pro.com/blog/p/14545-signature-sounds-explained-guitar-pro-7}}.

The data was first preprocessed using Music2Latent into individual four-second audio chunks prior to training. The model was then trained on a NVIDIA RTX A5000 GPU for 50 epochs (3,900 total steps) with a batch size of 64 and learning rate of $0.0001$, which took roughly 12 minutes in total. While this is a surprisingly small number of training steps, we found it to be sufficient for convergence.

\subsection{Evaluation}
Both the Guitar Pro and GuitarFlow audio are evaluated against the real DI audio in the test split of GOAT. Because electric guitar is almost always heard through the distortion of a guitar amplifier (whether subtle or heavy distortion) rather than the pure DI, all of the evaluation audio was separately rendered using a digital guitar amplifier\footnote{The ``crunch'' amplifier and default cabinet IR were used from \href{https://neuraldsp.com/plugins/archetype-nolly}{https://neuraldsp.com/plugins/archetype-nolly}} plugin. We then evaluate both the DI and amplifier conditions of the evaluation audio. Through this, we test two hypotheses: $\bm{H_1}$: audio from GuitarFlow sounds more realistic than the original Guitar Pro rendered audio as DI; and $\bm{H_2}$: audio from GuitarFlow sounds more realistic than the original Guitar Pro rendered audio when rendered using a distorted guitar amplifier.

Fréchet Audio Distance (FAD) \cite{kilgour2018fr} is a commonly used metric to measure audio similarity, while Kernel Audio Distance (KAD) \cite{kad} was recently proposed as a more flexible alternative which does not rely on a normality assumption of the audio embeddings. We calculate both FAD and KAD to evaluate the closeness of the synthetic and transformed audio to the real guitar DI recordings, using the real audio as the parent distribution. Several different embeddings \cite{hershey2017cnn} \cite{CLAP2022} \cite{kong2020panns} \cite{defossez2022highfi} \cite{cramer2019look} are used in these calculations. Following Hawthorne et al. \cite{hawthorne2022multi}, we also calculate the reconstruction embedding distance between the real DI audio and both the Guitar Pro audio and GuitarFlow output audio. This gives a better metric of how closely the actual audio is to the intended target, unlike FAD and KAD which work over a distribution of audio embeddings. This distance is obtained by calculating the Frobenius norm for both embeddings, and averaged over all time frames.

A simple listening test was also conducted with 16 participants (13 male, 3 female, with an average age of 27.8 years) to evaluate the model outputs. We selected 15 four-second outputs which covered single notes, chords, and playing techniques such as bends and muted strings. Participants were asked to rate each in terms of realism, which we define as how close the audio resembles a human playing guitar. All audio examples were normalised to -9dB RMS prior to the amplifier in order to ensure that the gain staging was consistent. Participants were paid with a £10 Amazon voucher. %Both the DI and amplifier variations of the stimuli were used. In order to not bias the selected examples while still retaining a large diversity in sound, the outputs from the testing split were divided into 15 separate categories, each representing specific notes or playing techniques. One example was then randomly selected from each category.

Finding a baseline model to compare against is difficult as most synthesis models use MIDI as a input sequence which cannot replicate any of the numerous expressive techniques, creating an unfair comparison with GuitarFlow. Many common style transfer approaches are also difficult to compare due to monophonic constraints \cite{caillon_rave_2021} \cite{engel_ddsp_2020} or large computational and data requirements \cite{huang2024musical} \cite{comanducci2023timbre}. Since our primary goal in this work is to establish the feasibility of flow-matching-based transfer in the context of symbolic-to-real audio synthesis, we focus on evaluating the improvement that the model makes compared to the intermediary synthetic audio within that pipeline. We leave a comprehensive comparison with other style transfer methods as an important direction for future work.

\section{Results and Discussion}

%A test split containing three songs (five percent of the dataset) which are completely unseen in the training data was used to prevent bias in the test results. 

\subsection{Objective Metrics}
\label{sec:obj_metrics}

%The FAD, KAD and reconstruction distances were all calculated on the testing split of the dataset and are presented in Table \ref{tab:results}. In general, GuitarFlow was able to achieve lower FAD and KAD when compared to the synthetic Guitar Pro audio. This is an indication that the features generated by GuitarFlow are more similar to the real DI audio than the synthetic audio. Of all the embeddings used, only VGGish resulted in a lower FAD for the synthetic Guitar Pro audio. However, given that both the FAD and KAD scores using the VGGish embedding are extremely similar, it is possible that this embedding is simply not well suited for representing the somewhat nuanced difference between a real guitar and a virtual instrument. This seems to follow findings that embeddings trained for classification such as VGGish are not well suited for FAD \cite{gui2024adapting}. GuitarFlow also achieved lower reconstruction distance scores on three of the five embeddings, indicating that it likely successfully transforming the audio to be closer to the intended target.

The FAD, KAD and reconstruction distances are presented in Table \ref{tab:results}. GuitarFlow generally performs much better than Guitar Pro, particularly in the DI condition. The amplifier condition is a bit more mixed, with GuitarFlow struggling to improve on Guitar Pro in the KAD metric. However, the FAD and reconstruction distance results are still strong in the amplifier condition. It is possible that the clipping and distortion from the amplifier removed some information from the DI which affected the calculated embeddings in a way that the KAD metric is sensitive to.

% NEW TABLE WITH RESULTS FROM LARGE TEST SPLIT
\begin{table*}[t]
    \centering
    \caption{Comparison of the original synthetic Guitar Pro virtual instrument (GP) and GuitarFlow. We calculate Fréchet Audio Distance (FAD), Kernel Audio Distance (KAD) and reconstruction distance (Recon. Dist.) on both the DI and the DI rendered through a guitar amplfier. Best values for each metric in each row marked in bold.}
    \label{tab:results}
    \begin{adjustbox}
    {width=\textwidth}   
    \begin{tabular}{llS[table-format=3.4]S[table-format=3.4]S[table-format=3.4]S[table-format=3.4]S[table-format=3.4]S[table-format=3.4]}
        \toprule
        \multirow{2}{*}{Condition} & \multirow{2}{*}{Embedding Model} & \multicolumn{2}{c}{FAD $\downarrow$} & \multicolumn{2}{c}{KAD $\downarrow$} & \multicolumn{2}{c}{Recon. Dist. $\downarrow$} \\
        %\cmidrule(lr){3-4} \cmidrule(lr){5-6} \cmidrule(lr){7-8}
        & & GP & GuitarFlow & GP & GuitarFlow & GP & GuitarFlow \\
        \midrule
        \multirow{5}{*}{DI} 
        & VGGish  & 2.7121458053588867 & \bfseries 2.3531885147094727 & 8.159232139587402 & \bfseries 5.71819543838501 & \bfseries 0.7890625 & 0.90380859375 \\
        & CLAP    & 207.2625732421875 & \bfseries 120.73974609375 & 17.65347671508789 & \bfseries 4.651451110839844 & 0.09014892578125 & \bfseries 0.08148193359375 \\
        & PANNs   & 16.830615997314453 & \bfseries 8.484001159667969 & 17.304336547851562 & \bfseries 4.836428165435791 & 0.96142578125 & \bfseries 0.65869140625 \\
        & EnCodec & 39.118858337402344 & \bfseries 14.737373352050781 & 22.887479782104492 & \bfseries 9.698325157165527 & 1.818359375 & \bfseries 1.396484375 \\
        & OpenL3  & 64.71243286132812 & \bfseries 37.97874450683594 & 8.204030990600586 & \bfseries 3.334057331085205 & \bfseries 1.5615234375 & 1.720703125 \\
        \midrule
        \multirow{5}{*}{Amplifier} 
        & VGGish  & 1.9607410430908203 & \bfseries 0.7668571472167969 &  \bfseries4.169094562530518 & 5.278873443603516 & \bfseries 1.087890625 & 1.1005859375 \\
        & CLAP    & 80.25076293945312 & \bfseries 33.40521240234375 & 8.731853485107422 & \bfseries 8.612465858459473 & 0.04620361328125 & \bfseries 0.036834716796875 \\
        & PANNs   & 13.899513244628906 & \bfseries 5.182134628295898 & \bfseries 6.5116286277771 & 9.985733032226562 & 0.99853515625 & \bfseries 0.73779296875 \\
        & EnCodec & 11.757339477539062 & \bfseries 4.392913818359375 & \bfseries 8.196043968200684 & 13.767873764038086 & 1.5673828125 & \bfseries 0.93115234375 \\
        & OpenL3  & 48.930816650390625 & \bfseries 20.615997314453125 & 5.286264419555664 & \bfseries 5.060231685638428 & 1.767578125 & \bfseries 1.41796875 \\
        \bottomrule
    \end{tabular}
    \end{adjustbox}

\end{table*}

\subsection{Listening Test}

\begin{figure}
    \centering
    \includegraphics[width=0.6\textwidth]{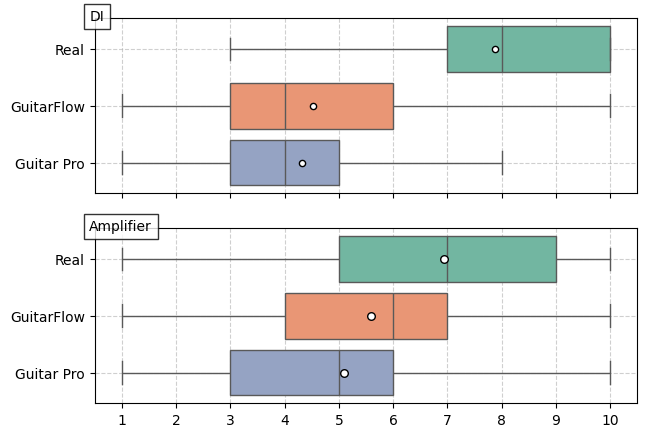}
    \caption{Boxpot with mean indicators for the MOS results of both the DI and amplifier conditions of the listening test. Mean scores in each group marked by white dots.}
    \label{fig:listeningtest}
\end{figure}

The listening test mean opinion score (MOS) results are shown in Figure \ref{fig:listeningtest}. Friedman tests revealed significant differences between groups in both the DI ($\chi^2(3) = 267.114$, $p < .001$) and amplifier ($\chi^2(3) = 107.153$, $p < .001$) conditions. A pairwise Wilcoxon signed-rank test was then performed for both conditions using a Bonferroni-corrected $\alpha=0.0167$. For the DI stimuli, significant differences were found between the real and GP $(p<.001)$ groups and the real and GuitarFlow $(p<.001)$ groups. For the amplifier stimuli, significant differences were found between the real and GP $(p<.001)$ groups, the real and GuitarFlow $(p<.001)$ groups and the the GP and GuitarFlow $(p<.001)$ groups.

While the GuitarFlow model is only barely perceived as more realistic than the Guitar Pro audio in the DI scenario ($\bm{H_1}$), it is perceived as significantly more realistic when run through the distortion of a guitar amplifier ($\bm{H_2}$). It is interesting that the realism MOS would increase after the distortion is applied to the signal, as this distortion clips the signal and loses information. This also seems to contrast the KAD results of the amplifier condition, though this discrepancy could be due to a limitation of the KAD metric, the embeddings used, or simply a difference between what each result is measuring. We also see the MOS scores of the Guitar Pro and GuitarFlow stimuli increase and the real stimuli decrease in the amplifier condition. In their post-survey remarks, one participant noted that some of the examples had ``characteristic sounds of neural synthesis''. We theorise that this amplifier distortion process helps the participants to focus more on the realism of the example, in the context of its content and timbre. This is because the amplifier distortion process hides the aforementioned neural artifacts that would otherwise cause the MOS to be lowered, as in the DI case, and distract listeners from the task at hand. %Furthermore, this process renders the samples closer to an ecologically valid scenario, as it is not common to listen to and assess DI recordings of guitar.

\subsection{Subjective Analysis}\label{subj-anal}

A careful subjective listening analysis of the audio examples revealed that GuitarFlow seems to particularly excel at recreating strumming chords. This is particularly relevant given that most commercially available guitar virtual instrument software notoriously struggle with this technique. However, the model appears to struggle much more when generating single notes, creating obvious neural artifacts which are not present during strummed chords. Due to this discrepancy, we hypothesize that this could be due to the notes of the real guitar DI not being perfectly aligned to the notes in the Guitar Pro rendering. As chords are strummed, they inherently have a looser timing window compared to single notes. This should be addressed in a future listening study which focuses on strumming vs. single notes. %We theorize that one possible method to investigate this could be through aligning the MIDI notes of the annotation and rendering it with a MIDI virtual instrument instead of Guitar Pro. 

%A careful subjective listening analysis of the audio examples revealed that GuitarFlow seems to particularly excel at recreating strumming chords. This is particularly relevant given that most commercially available guitar virtual instrument software notoriously struggle with this technique. However, the model appears to struggle much more when generating single notes, creating obvious neural artifacts which are not present during strummed chords. Due to this discrepancies, we hypothesize that this could be due to the notes of the real guitar DI not being perfectly aligned to the notes in the Guitar Pro rendering. This happens because Guitar Pro assumes notes to be played perfectly to a tempo grid, while real audio contains small timing imperfections due to being played by a human. As chords are strummed, they inherently have a looser timing window compared to single notes. We theorize that one possible method to investigate this could be through aligning the MIDI notes of the annotation and rendering it with a MIDI virtual instrument instead of Guitar Pro. 

\subsection{Limitations \& Future Work}

As in many deep learning related works, data is the main limiting factor in our approach. While the experiment showed great promise despite its data constraints, additional data points from a more varied pool of guitars and guitarists could potentially allow for better sound quality and generalisability, as well as explicit style controllability. Unfortunately, obtaining paired examples of guitar audio and tablatures is an expensive and time consuming process. Pretraining on synthetic data or using unpaired training methods may help address this. The annotation alignment issue may have also adversely affected the final audio quality. However, the experiment undertaken alone is not sufficient to clarify this hypothesis. Additionally, the listening test only measures a broad ``realism" of the synthesis quality, and thus we are unable to gain any insight to any more fine-grained aspects of the results such as the timbre, note accuracy or human-like variability.
\section{Conclusion}

In this paper we presented GuitarFlow, a novel methodology and model for synthesising realistic electric guitar from guitar tablatures. This model makes use of latent Flow Matching to perform style transfer on a basic audio render of the tablature, allowing the model to be trained quickly on an extremely small amount of data while still generalising to unseen data. This approach was justified through several objective metrics and a listening test. We hope that our work will inspire more researchers to investigate Flow Matching for generative audio, as well as work on generative systems which allow for increased musical control.

\begin{credits}
\subsubsection{\ackname} 
This work is supported by the EPSRC UKRI Centre for Doctoral Training in Artificial Intelligence and Music (Grant no. EP/S022694/1) and UKRI - Innovate UK (Project no. 10102804).

\subsubsection{\discintname}
The authors have no competing interests.
\end{credits}
 \bibliographystyle{splncs04}
 \bibliography{ref}

%\end{thebibliography}
\end{document}